\documentstyle[prc,aps,psfig]{revtex}
\begin{document}

\draft

\title{Nuclear Reaction Rates and Primordial $^6$Li}

\author{Kenneth M. Nollett$^{1}$, Martin Lemoine$^{2,3}$, and David N. 
Schramm$^{1,2,3}$}
\address{
$^1$ Department of Physics, Enrico Fermi Institute, The University 
of Chicago, Chicago, IL 60637-1433\\}
\address{
$^2$ NASA/Fermilab Astrophysics Center, Fermi National Accelerator 
Laboratory, Batavia, IL 60510-0500 \\}
\address{
$^3$ Department of Astronomy and Astrophysics, Enrico Fermi 
Institute, The University of Chicago, Chicago IL  60637--1433}

\date{\today}

\maketitle

\begin{abstract}
We examine the possibility that
Big Bang Nucleosynthesis (BBN) may produce non-trivial amounts
of $^{6}{\rm Li}$.  If a primordial component of this isotope
could be observed, it would provide a new fundamental test of
Big-Bang cosmology, as well as new constraints on the baryon
density of the universe.  At present, however, theoretical
predictions of the primordial $^{6}{\rm Li}$ abundance are
extremely uncertain due to difficulties in both theoretical
estimates and experimental determinations of the $^{2}{\rm
H}(\alpha,\gamma)^{6}{\rm Li}$ radiative capture reaction
cross-section. We also argue that present observational
capabilities do not yet allow the detection of primeval $^6$Li
in very metal-poor stars of the galactic halo.  However, if the
critical cross section is towards the upper end of its
plausible range, then improvements in $^{6}{\rm Li}$ detection
capabilities may allow the establishment of $^{6}{\rm Li}$ as
another product of BBN.  It is also noted that a primordial
$^{6}{\rm Li}$ detection could help resolve current concerns
about the extragalactic D/H determination.
\end{abstract}


\section{Introduction}

The consistency of the observed light element abundances with the
predictions of Big-Bang nucleosynthesis (BBN) is a fundamental source of
evidence for a hot Big-Bang\cite{peebles}.  Over the last thirty years, the
abundances of the light isotopes $^2$H, $^{3}{\rm He}$, $^{4}{\rm
He}$, and $^{7}{\rm Li}$ have all been found to be consistent with the
primordial levels predicted by BBN over a fairly narrow range of the
baryon-to-photon ratio of the universe, $\eta$: $2.5\times 10^{-10} <
\eta < 6\times 10^{-10}$ (see, {\it e.g.} Ref.\cite{copi} and references
therein).
The fact that there is such a range of concordance, for abundances 
spanning more than nine orders of magnitude, is taken as evidence that
BBN gives a correct description of the origin of the light elements. 
This concordance interval also provides a measure of the baryonic
contribution to the total mass density of the universe, 
$0.01h^{-2} < \Omega_B < 0.02h^{-2}$, as obtained from the constraints 
on $\eta$, and where $h$ denotes the value of the Hubble constant in 
units of 100 km/s/Mpc.

\begin{figure}
\hskip .25in
\centerline{\psfig{file=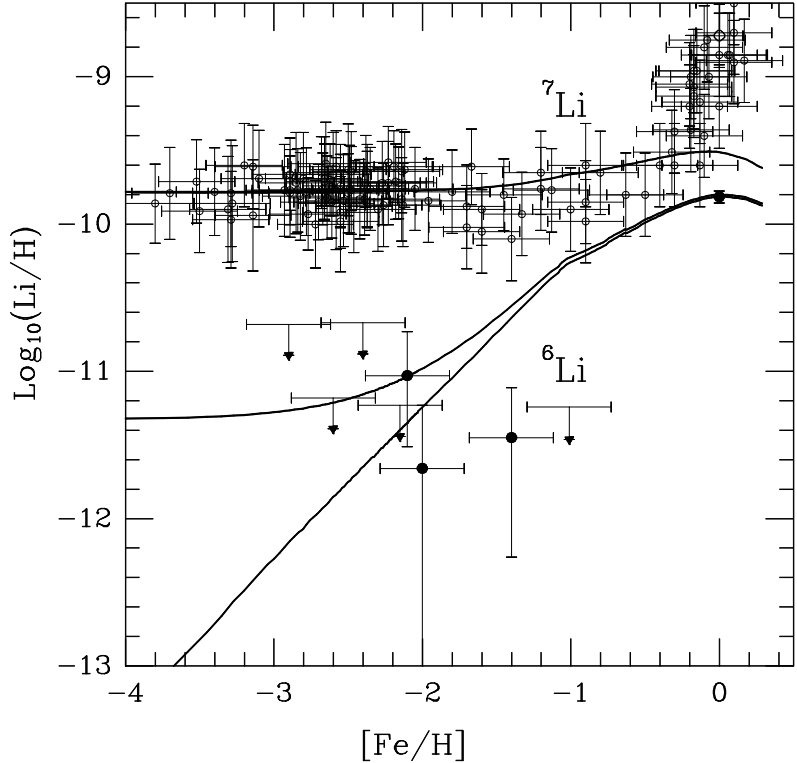,width=5in}}
\caption{Abundance {\it vs.} Metallicity.  Open data points represent $^7$Li
abundances, the flat region at low metallicities being attributed to the 
primordial abundance of $^7$Li. Filled data points represent $^6$Li abundance
measurements and triangles refer to upper limits. The solid curves bracket the
possible
primordial abundances of $^6$Li and indicate the evolution of $^6$Li with
metallicity, assuming $^6$Li evolves like $^9$Be (see 
Ref.{\protect\cite{martin}}). A
primordial component of $^6$Li would show up as a flat region of the curve at
low metallicity, as shown here for the upper limit derived from 
Ref.{\protect\cite{cecil}}, ($^6$Li/H)$_{BBN}\sim5.\times10^{-12}$.
The lower limit corresponds to ($^6$Li/H)$_{BBN}\sim10^{-14}$.}
\label{fig:evolution}
\end{figure}

Inferring primordial abundances of elements is a tricky business, and
it seems fair to say that at the present time, the constraints on the
baryon density are limited by the systematic errors on the observed or
inferred abundances \cite{copi}. In this regard, an additional light
isotope could further firm up BBN, and might provide new constraints
on $\eta$.  The only remaining candidate that could in principle be
brought into the framework of homogeneous BBN is $^{6}{\rm Li}$.
$^{6}{\rm Li}$ has the next-highest predicted primordial abundance,
after those species already understood in the BBN framework.  (See
Ref. \cite{wagoner}.)  Like beryllium and boron, present day $^{6}{\rm
Li}$ is thought to be produced mostly by cosmic ray spallation in the
galaxy \cite{rfh}.  The meteoritic abundance of $^{6}{\rm Li}$ is
certainly much higher (by a factor $\sim 100$) than even the most
optimistic primordial abundance predicted by standard BBN.  However,
it is possible that the levels of this isotope in hot
$T_{eff}\sim$6000-6300 K, extreme low-metallicity halo stars ( either
main-sequence dwarfs or subgiants near the turn-off point) could
reflect its primordial abundance.  This would show up as a flattening
of the curve of $^{6}{\rm Li}$ {\em vs.} metallicity at the point
where the abundance of cosmic-ray-produced $^{6}{\rm Li}$ becomes
comparable to the abundance of primordial $^{6}{\rm Li}$ (see
Fig.\ref{fig:evolution}).  
Such a situation appears to hold for
$^{7}{\rm Li}$, whose abundances in low-metallicity halo stars are
uniform over a wide range of metallicities and a narrow range of
temperatures, the so-called Spite plateau
\cite{spite,thorburn,spite2}.

At present, there have been only three relatively uncertain detections
of $^6$Li in such low-metallicity stars, one of them being marginal
\cite{nissen}. The metallicities of these stars where $^6$Li has been
observed, roughly [Fe/H]$>-2.$\footnote{[Fe/H]=log$_{10}\left({\rm
Fe/H}\right) - {\rm log}_{10}\left({\rm Fe/H}\right)_{\odot}$, where
the subscript $\odot$ refers to abundances measured at the birth of
the Sun}, are unfortunately not low enough for any primordial component
to be observable. However, as new data come in, and as new instruments
that are able to reach lower metallicities and lower $^{6}{\rm Li}$
abundance levels eventually come on line, it is of interest to know
what levels of primordial $^{6}{\rm Li}$ we might expect to see, and
to what extent they could provide constraints on the baryon density.

Thomas {\it et al.} \cite{thomas} have examined the BBN predictions for the
primordial abundance of $^6$Li. These authors have not discussed in
detail, however, the extremely large uncertainties on this prediction,
as they were mainly concerned with both homogeneous and
non-homogeneous nucleosynthesis yields of beryllium and
boron. Predictions of primordial abundances are made by numerically
integrating rate equations for nuclear reactions that occurred during
the first few minutes of the Big Bang, and for $^6$Li, the
uncertainties on the yields are directly related to uncertainties on
the input reaction rates. Therefore, we first examine the status of
relevant cross-section measurements and identify the chief sources of
uncertainty.  We then discuss the prediction of the $^6$Li primordial
abundance, and we examine to what extent primordial $^6$Li could be
observed. We argue that even in the most optimistic case, this
observation is not within reach of present instrumental capabilities,
but must be subjected to future techniques. In particular, we discuss
how a direct measurement of the $^{2}{\rm H}(\alpha,\gamma)^{6}{\rm
Li}$ radiative capture cross-section, at the low energies where this
reaction takes place during BBN, $E\sim$60 -- 400 keV, could have a
profound impact on the predictions.  In fact, the present uncertainty
on the $^6$Li yield is so large that even if $^6$Li were detected in
very metal-poor stars, at metallicities of about [Fe/H]$<-3$, this
would not allow a sensible constraint on the baryonic density
parameter. However, an eventual measurement of the primordial $^6$Li
abundance, at a predicted level ($^6$Li/H)$\sim10^{-14}-10^{-12}$,
would in any case provide another fundamental test of modern
cosmology.

\section{Reaction rates}

The primordial abundance of $^{6}{\rm Li}$ is determined almost
entirely by the rates of two reactions.  These reactions are radiative
capture of deuterium on alpha particles, $^{2}{\rm
H}(\alpha,\gamma)^{6}{\rm Li}$, which produces practically all of the
$^{6}{\rm Li}$, and the $^{6}{\rm Li}$ {\em destroying}
reaction $^{6}{\rm Li}({\rm p},\alpha)^{3}{\rm He}$.  We examine below
the current status of these reaction rates.

\begin{figure}
\centering
\centerline{\psfig{file=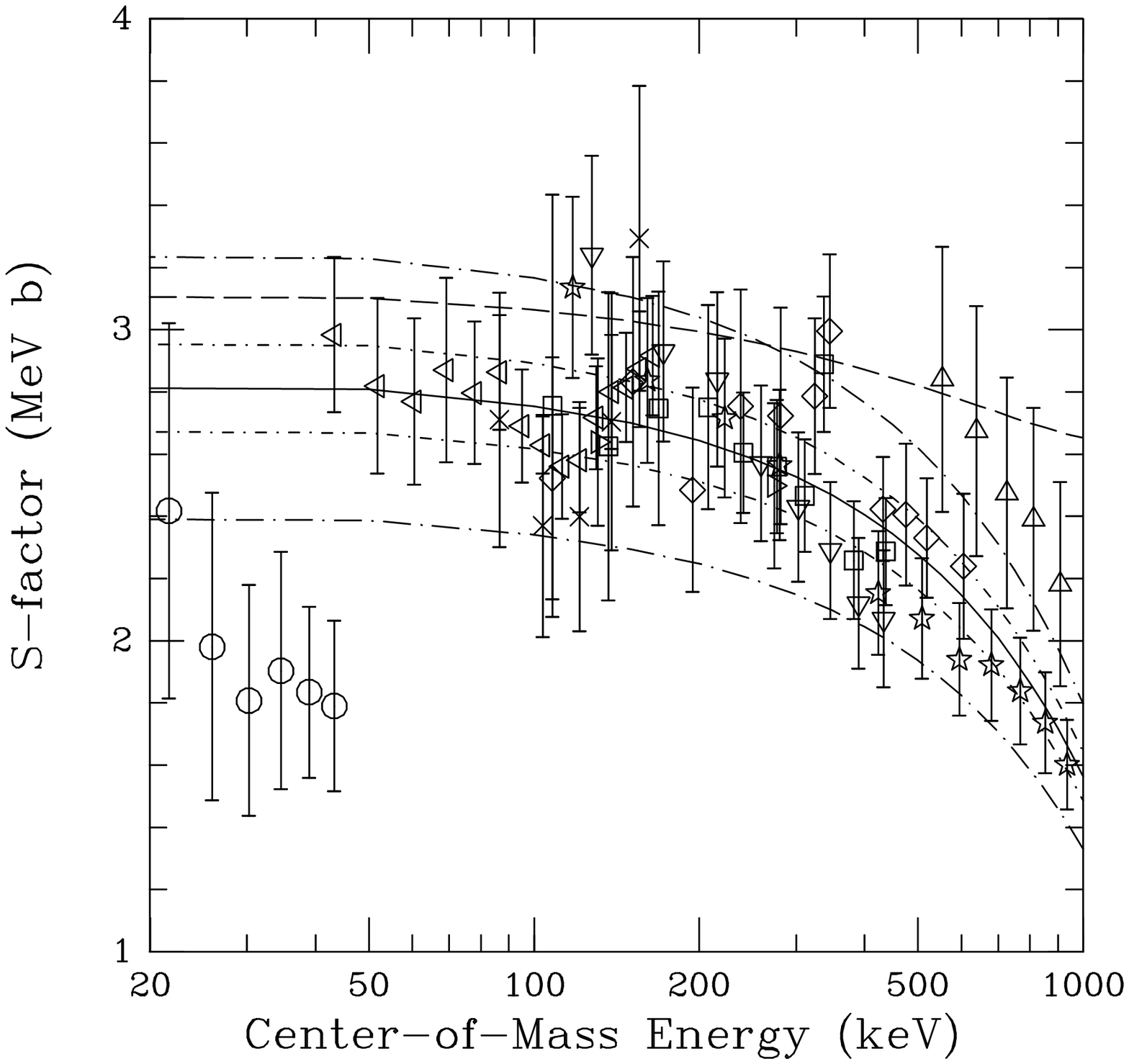,width=5in}}
\vskip .75in
\caption{$^{6}{\rm Li}({\rm p},\alpha)^{3}{\rm He}$ S-factor.
Experimental data for the reaction are shown, along with our fit
(solid line),
``1$\sigma$'' and ``2$\sigma$'' uncertainties in our fit
(symmetrical dot-dash lines), and the
standard fit of Harris {\it et al.}{\protect \cite{FCZ3}} (long
dash line).  Data are those of 
$\rhd$ Gemeinhardt {\protect\cite{gemeinhardt}},
$\bigcirc$ Fiedler \& Kunze {\protect\cite{fiedler}},
$\lhd$ Spinka {\it et al.} {\protect\cite{spinka}},
$\Box$ Kwon {\it et al.} {\protect\cite{kwon}}, 
$\Diamond$ Shinozuka {\it et al.} {\protect\cite{shinozuka}},
$\star$ Elwyn {\it et al.} {\protect\cite{elwyn}},
$\bigtriangleup$ Marion {\it et al.} {\protect\cite{marion}},
$\times$ V\'{a}rnagy {\it et al.} {\protect\cite{varnagy}}, 
and
$\bigtriangledown$ Engstler {\it et al.}{\protect\cite{engstler}}. 
}
\label{fig:p-a}
\end{figure}

\subsection{The reaction ${\bf ^{6}{\rm Li}({\rm p},\alpha)^{3}{\rm He}}$}
\label{sec:p-alpha}

The low-energy (100 keV $< E < $1000 keV) cross-section for this
reaction is sufficiently well-known that recent work \cite{engstler}
has concentrated on determining the effects of electron screening in
the experimental target at extremely low energies ($E < 100$ keV) via
comparison with the higher-energy cross section.  The energy range
that concerns us here is the range in which the peaks of the Coulomb
barrier penetration factor and of the Maxwell-Boltzmann
thermal velocity distribution overlap significantly at BBN
temperatures.  It is in this range, where there is a population of
protons with enough thermal energy to penetrate the Coulomb barriers
of the $^{6}$Li ions, that the reaction takes place.  (See Ref. \cite{rnr}
for a detailed discussion.)  In the case of $^{6}{\rm Li}({\rm
p},\alpha)^{3}{\rm He}$, this corresponds to energies of $E \sim$ 75 --
410 keV at a temperature of $10^{9}$K at the beginning of BBN
or $ E \sim$ 30 -- 80 keV at a temperature of $10^{8}$K, when the
$^6$Li abundance has stabilized.

For purposes of fitting curves to experimental cross section
data and integrating them to obtain reaction rates, it is
customary to use the astrophysical S-factor, defined by removing
the Coulomb barrier factor and a geometric factor from the cross section:
\begin{equation}
S(E) = E\sigma(E)\exp\left[-\left(E_{g}/E\right)^{1/2}\right],
\end{equation}
where $E$ is energy, $\sigma$ is the reaction cross section, and
and $E_{g}$ is the Gamow energy,
\begin{equation}
E_{g} = 2\mu\pi^{2} e^{4}Z_{1}^{2}Z_{2}^{2}/\hbar^{2},
\end{equation}
for reactants of reduced mass $\mu$ and atomic number $Z_{1}$ and
$Z_{2}$. (See Ref. \cite{rnr}.)  The S-factor is particularly
convenient for fitting because it is often a much slower function of
$E$ than the cross section is.  (For the procedure used to derive a
reaction rate from the astrophysical S-factor, see Ref.\cite{rnr}.)
We have computed a new analytic expression for the $^{6}{\rm Li}({\rm
p},\alpha)^{3}{\rm He}$ reaction rate using a new polynomial fit to
the experimental S-factor between 100 and 1000 keV.  (See Table
\ref{table:rates} and Fig. \ref{fig:p-a}.)  
In addition, we present
the S-factor curve corresponding to the rate found in the compilation
of Harris {\it et al.}  \cite{FCZ3}.  Following Engstler {\it et
al.}\cite{engstler}, we use only the data above 100 keV
\cite{gemeinhardt,kwon,fiedler,spinka,shinozuka,elwyn,marion,varnagy,%
engstler} in the fit to avoid the effects of electron screening.
Unlike their fit, ours includes their data in addition to previous
data.  Our reaction rate is lower than that of Harris {\it et al.} by
a factor of about 15\%.  Treating all errors as statistical in our
least-squares fit to the cross-section data gives a $1\sigma$ error of
5\% in overall normalization (based on the fitting error at 100 keV,
an energy relevant to BBN).  An estimated $2\sigma$ error of 15\%
includes all of the lowest data points except those of Fiedler and
Kunze \cite{fiedler}, which were not used in the fit, and which seem
to be normalized differently from the rest of the data.  We will use
this rather extreme estimate to determine upper limits on the
$^{6}{\rm Li}$ yield.  From this estimate, we still find the
uncertainty in this reaction rate to be insignificant in comparison to
uncertainties in the main $^{6}{\rm Li}$-producing reaction rate.
(See below.)

\subsection{The reaction ${\bf ^{2}{\rm H}(\alpha,\gamma)^{6}{\rm Li}}$}

In contrast, the low-energy cross section for radiative capture of a
deuteron by an alpha particle to form $^{6}{\rm Li}$ is almost
completely unknown.  Theoretical calculations (refs.
\cite{langanke1,langanke2,typel,rghr,mohr,ryzhikh,alma-ata,crespo})
vary over a factor of about ten.  Experimental measurements are
difficult because of the extremely small cross sections involved;
electric dipole radiation is strongly suppressed because the nearly
equal charge-to-mass ratios of the deuteron and alpha particle give
the $d + \alpha$ system a very small dipole moment in all cases.  This
requires the radiative capture to proceed mostly via electric
quadrupole radiation, and thus mostly through the $d$-wave portion of
the incoming wavefunction.

\begin{figure}
\centering
\centerline{\psfig{file=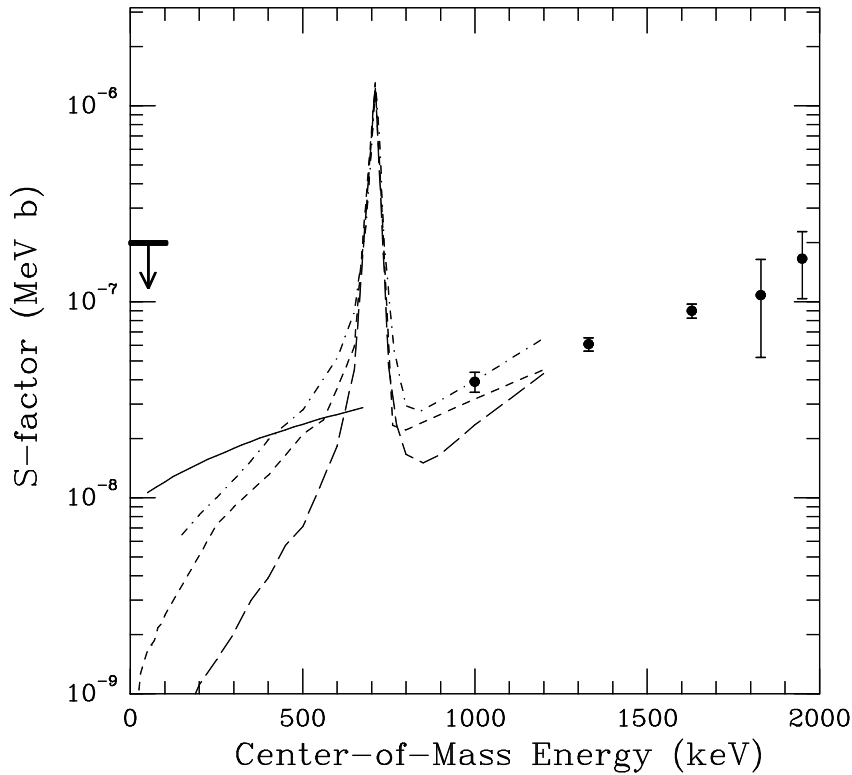,width=5in}}
\vskip .3in
\caption{ d--$\alpha$ Capture S-factor.  A selection of measured and
inferred astrophysical S-factors for the reaction $^{2}{\rm
H}(\alpha,\gamma)^{6}{\rm Li}$ is shown.  In order of decreasing
low-energy S-factor, the calculations are from Coulomb-breakup
measurements of Kiener {\it et al.}{\protect\cite{kiener}}, and
the models of Mohr {\it et al.}{\protect\cite{mohr}}, Ryzhikh 
{\it et al.}{\protect\cite{ryzhikh}}, and Typel
{\protect\cite{typel}}.
The 53 keV upper limit is from Cecil {\it et al.}{\protect\cite{cecil}}, and
all other points are from Robertson {\it et
al.}{\protect\cite{rghr}}.
The data of Mohr {\it et al.} (not shown) are concentrated at
the top of the 711 keV resonance.}
\label{fig:d-a}
\end{figure}

To date, there have been three experiments to directly measure
cross-sections for $d + \alpha$ radiative capture.  The only recent
direct measurement of the nonresonant cross section, used in the
current standard low-energy extrapolation \cite{CF}, is that of
Robertson {\it et al.} \cite{rghr}, who measured the reaction cross
section at center-of-mass energies of 1--3.5 MeV.  The experiment of
Mohr {\it et al.} \cite{mohr} concentrated on the $J^{\pi} = 3^{+}$
resonance at 711 MeV.  By contrast, the energies relevant to $^{6}{\rm
Li}$ production in the big bang are in the range 30 -- 400 keV.  The
recent experiment of Cecil {\it et al.} \cite{cecil}, has determined
an upper limit (at 90\% confidence level) for the cross section at 53
keV.  Unfortunately, this limit is much higher than any current
theoretical estimate of the cross section at 53 keV, so the actual
reaction rate may be much lower -- by a factor of fifty or so -- than
the limit implied by this measurement.  (See Fig.\ref{fig:d-a}.)

In recent years, an attempt has been made to get around the difficulty
of measuring very small radiative-capture cross sections by studying
Coulomb breakup of the product nucleus as an inverse reaction
\cite{baur,kiener,hesselbarth}.  In this scheme, $^{6}{\rm Li}$ nuclei
are Rutherford-scattered off some high-Z nucleus.  Some of these
scattered nuclei are broken up into a deuterons and alpha particles by
the electric-field gradient of the heavy nucleus.  This process has
been treated as the absorption of a
virtual photon, and thus as an inverse radiative capture \cite{baur}.
However, a number of difficulties arise in treating the data
\cite{cecil,gazes} which produce additional uncertainties in this
approach, mostly because contributions from the various partial waves
are not the same in Coulomb breakup as they are in radiative capture.
One group \cite{hesselbarth} reports anomalous angular dependence in
the data.  The cross sections inferred from the breakup measurements
of Kiener {\it et al.} \cite{kiener} are significantly higher than
any of the theoretical estimates, perhaps suggesting interference
from the nuclear force, even at small scattering angles, or perhaps
supporting a higher than anticipated low-energy cross section.

It is clear from the conflicting theoretical curves that a reliable
determination of the reaction rate for $d + \alpha$ radiative capture
will require a direct cross section measurement below the $J^{\pi} =
3^{+}$ resonance.  While the cross section was too small to be
measured at 53 keV (an alpha particle energy of 160 keV), the expected
cross section should exceed this limit slightly higher in the range of
energies $\sim$50 -- 400 keV relevant to BBN (alpha bombarding
energies of $\sim$150 -- 1200 keV) so that it can be measured in
similar experiments.  In the absence of any evidence to allow a
decision between the various theoretical and experimental
extrapolations, we will use a few representative results to see how
large a spread is allowed.  It should be kept in mind that these
cross-sections are only representative values selected from the
literature, where it is difficult to choose between conflicting
results.  (Note, for example, that in the same paper where they
calculate one of the cross sections used here, Ryzhikh {\it et al.}
\cite{ryzhikh} also obtain significantly different results by
expanding the $^{6}{\rm Li}$ ground state in different sets of basis
functions.)  We will also use the Cecil {\it et al.}\cite{cecil} value
as an extreme upper limit, setting the nonresonant S-factor constant
at the 53 keV limit of $2\times10^{-7}$ MeV$\cdot$b for energies below
the 711 keV resonance.

Using the measurement of the resonant cross section due to Mohr {\it
et al.}, \cite{mohr}, we also present a new value of the contribution
to the reaction rate from the 711 keV resonance.  Using the methods
described in \cite{rnr}, the resonant contribution is
\begin{equation}
N_{A}\langle\sigma v\rangle_{\rm Resonant} = 97.1\/
T_{9}^{-3/2}\exp(-8.251/T_{9})
\end{equation}
This is a fairly small change from the customary value, given in
Robertson {\it et al.,}\cite{rghr} and it does not have a significant
effect on the rate at BBN temperatures or on the BBN $^{6}{\rm
Li}$ yields.

\begin{figure}
\centerline{\psfig{file=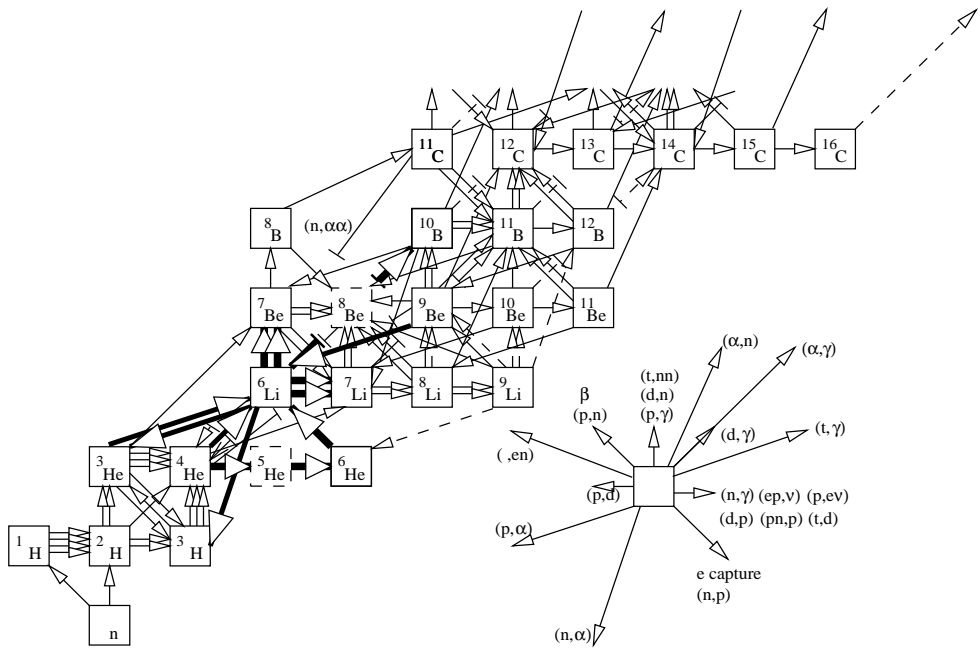,width=5.1in}}
\vskip .25in
\caption{BBN Reaction Network.  The lower portion of the BBN reaction
network used here, which is identical to that of Ref.
\protect\cite{thomas}.  Reactions producing or destroying $^6$Li
are indicated by thick lines.}
\label{fig:network}
\end{figure}

\subsection{Other $^6$Li reactions}

Eleven other reactions in the BBN reaction network \cite{thomas} involve
$^{6}{\rm Li}$ (see Table \ref{table:rates} and Fig.
\ref{fig:network}).  
However, neither removing these reactions
(individually) from the reaction network nor augmenting them by large
factors changes the final $^{6}{\rm Li}$ abundance by more than
one tenth of a percent.  The effect of any uncertainty in these rates
is certainly swamped by the uncertainties in the rates of the more
crucial reactions discussed above, so we pass over them.  Note, in
particular, the reaction $^{3}{\rm He}(^{3}{\rm H},\gamma)^{6}{\rm
Li}$, which has generally been omitted from BBN studies, has only a
very small effect on the $^{6}{\rm Li}$ yield \cite{fukugita}, so its
uncertainty was safely ignored.

\section{Primordial abundance ($^6$L\lowercase{i}/H)}

We predicted $^{6}{\rm Li}$ yields for various values of the relevant
reaction rates using Kawano's version \cite{kawano} of the standard
nucleosynthesis code and the full network of Thomas {\it et
al}\cite{thomas}.  We were particularly interested in establishing
upper limits for the primordial $^{6}{\rm Li}$ abundance to determine
whether it is possible in principle for primordial $^{6}{\rm Li}$ to
contribute a significant fraction of the $^{6}{\rm Li}$ abundance at
low metallicity.

\subsection{Predictions and uncertainties}

The uncertainty in the primordial abundance of $^{6}{\rm Li}$ depends
only weakly on the $^{6}{\rm Li}$-destroying proton capture reaction
considered in section \ref{sec:p-alpha} above.  Holding all other
rates at their standard values, a 20\% increase in this rate
decreases the $^{6}{\rm Li}$ yield by 10\%; a 20\% decrease in this
rate increases the $^{6}{\rm Li}$ yield by $\sim30$\%.  As
discussed above, the $2\sigma$ uncertainty in this reaction rate is
probably less than 20\%.  Therefore, uncertainties in this reaction
rate have at most a small role to play in determining the possibility
of observing primordial $^{6}{\rm Li}$.

\begin{figure}
\centering
\centerline{\psfig{file=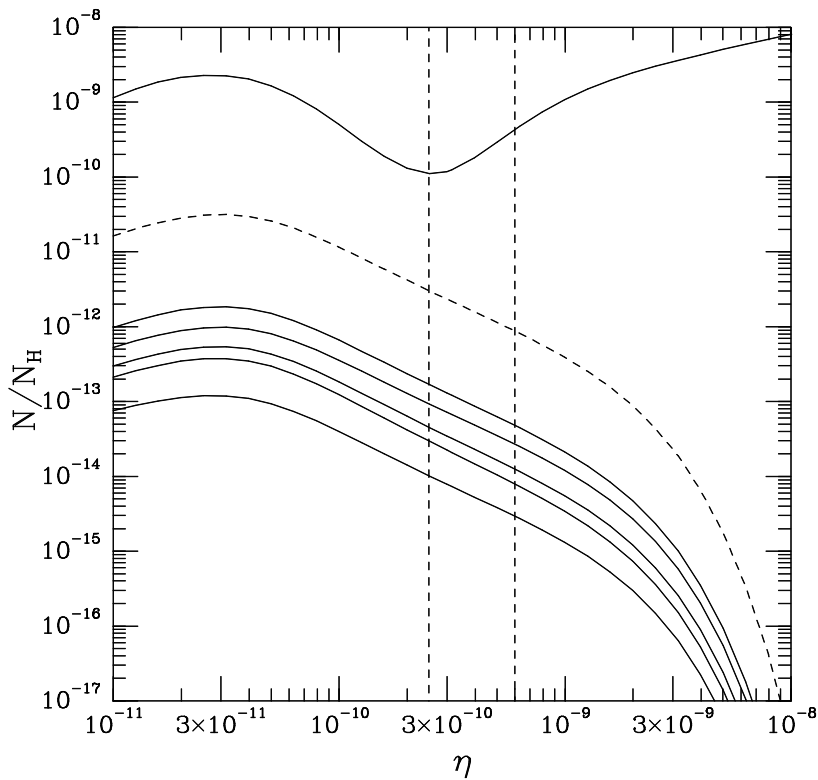,width=5in}}
\caption{{Predicted Abundances.}  Abundances relative to
hydrogen generated from the BBN network of figure
\protect\ref{fig:network}, with the concordance interval in
$\eta$ of Copi {\it et al.}\protect\cite{copi} indicated by
vertical dashed lines.  The top curve is the predicted $^{7}$Li
abundance.  The several other solid curves are results for
$^6$Li based on the Kiener {\it et al.}{\protect\cite{kiener}}
Coulomb-breakup measurements and the calculations of Mohr {\it
et al.}{\protect\cite{mohr}}, Robertson {\it et
al.}{\protect\cite{rghr}} (the ``standard'' rate), Ryzhikh {\it
et al.}{\protect\cite{ryzhikh}}, and Typel, {\it et al.}
{\protect\cite{typel}}, in decreasing order.  The dashed curve
is derived from the experimental limit of Cecil {\it et
al.}\protect\cite{cecil}.}
\label{fig:abundances}
\end{figure}

The dependence of the $^{6}{\rm Li}$ yield on normalization of the $d
+ \alpha$ radiative capture rate is very nearly linear at all values
of the normalization.  Given the wide range of predictions for the
reaction rates and the extreme upper limit from \cite{cecil}, there is
a wide range of possible $^{6}{\rm Li}$ yields.  Depending on the
low-energy extrapolation used, the maximum possible primordial ratio
of $^{6}{\rm Li}$ to $^{7}{\rm Li}$ varies from 0.01\% ($^{6}{\rm
Li}/{\rm H} = 1.4\times 10^{-14}$) to 0.18\% ($^{6}{\rm Li}/{\rm H} =
2.4\times 10^{-13}$), while the extreme upper limit on this ratio
derived from the Cecil {\it et al.} limit on the $d + \alpha$ cross
section and our lower limit on the $^{6}{\rm Li}(p,\alpha)^{3}{\rm He}$
cross section is as high as 3.7\%; see Fig. \ref{fig:abundances}.
The maximum always occurs at $ \eta \simeq 2\times 10^{-10}$, the
extreme low end of the concordance interval allowed by standard BBN
\cite{copi}. These results are in agreement with those of
Ref.\cite{thomas}, who found an upper limit on the primordial ratio
$^6$Li/$^7$Li$<0.2$\% in all cases they considered.

Because $^{6}{\rm Li}$ yields fall rapidly with increasing
baryon density, $^{6}{\rm Li}$ is potentially a very sensitive
probe of the baryon density.  However, a direct measurement of
the $^{2}{\rm H}(\alpha,\gamma)^{6}{\rm Li}$ reaction cross
section at low energy will be necessary before any such claims
can be made. In the meantime, it is obvious that a detection of
primeval $^6$Li at a level consistent with the above estimations
would provide a new piece of evidence for the consistency of
BBN, and hence a fundamental cosmological test.  While a precise
baryon density determination from $^{6}$Li is not possible
without a better rate for $^{2}{\rm H}(\alpha,\gamma)^{6}{\rm
Li}$, it is clear that a measureable primordial component of
$^{6}$Li would argue for a value of $\eta$ near the lower end of
the allowed range.  (See Fig. \ref{fig:abundances}.)  This is
also the range of $\eta$ implied by the extragalactic deuterium
measurements of Rugers and Hogan \cite{hogan}, but it would be
in conflict with the the lower D/H (higher $\eta$) values
implied by the work of Tytler {\it et al.} \cite{tytler}.

The highest value of the $^{6}{\rm Li}$/$^{7}{\rm Li}$ ratio may allow
something like the ``Spite plateau'' of $^{7}{\rm Li}$ to exist for
$^{6}{\rm Li}$ at extremely low metallicities. It was argued in
Ref.\cite{martin} that $^6$Li/H should scale as $^{16}$O/H all along
the galactic evolution, taking into account the trends observed for
$^9$Be.  This means that the curve $\log(^{6}{\rm Li}/{\rm H})$ {\it
vs.}  $\log({\rm Fe}/{\rm H})$ should have a slope unity in the halo
phase, {\it i.e.} up to [Fe/H]$\simeq-1.$, and a slope $\sim0$ during
the disk phase $-1<{\rm [Fe/H]}<0$ (see
Fig. \ref{fig:evolution}). Roughly speaking, since the meteoritic
abundance of $^6$Li is $\log(^{6}{\rm Li}/{\rm H})\sim-10.$ at
[Fe/H]$\equiv0$., one would expect the primordial $^6$Li plateau to
show up at [Fe/H]$\simeq-3$ if the primordial abundance is
$\log$($^6$Li/H)$_p\sim-12.$, at [Fe/H]$\sim-4.$ if
$\log$($^6$Li/H)$_p\sim-13.$, and so forth.  We note that the lower values
of the reaction rate would result in $^6$Li yields so much lower than
those expected from cosmic-ray spallation that they would not be
observable even in the least-evolved stars. On the other hand, the
extreme upper limit on the $d + \alpha$ cross section derived from the
experiment of \cite{cecil} corresponds to an upper limit of the
primordial $^{6}{\rm Li}$ abundance at about the level of previous
detections.

\subsection{Observing primordial $^6$Li}

It is extremely difficult to detect the absorption due to the presence
of $^6$Li in the photosphere of a metal-deficient star for the two
following reasons: {\it (i)} the only resonance line of $^6$Li{\sc i}
at 6708\AA~is usually blended with that of $^7$Li{\sc i} since the
isotopic separation is of the same order or smaller than the typical
width of the lines; {\it (ii)} $^6$Li is strongly under-abundant
compared to $^7$Li, especially at low metallicities where the
abundance of $^7$Li is constant, of the order of the meteoritic
abundance of $^6$Li, and the $^6$Li abundance goes down as the
metallicity.  The absorption of $^6$Li can therefore be seen only as a
slight asymmetry of the  $^7$Li absorption line profile.

Nonetheless, two detections of $^6$Li have probably been achieved at
metallicities [Fe/H]$\simeq-2.1$ and [Fe/H]$\simeq-1.4$, at a level
$^6$Li/$^7$Li$\simeq5$\% \cite{nissen}. The main limiting factors for
these detections were the signal-to-noise ratio and the spectral
resolution achieved by the instruments. However, the accuracy of the
measured value was limited equally by the noise in the observed
spectrum (statistical error) and by the accuracy of the determination
of the velocity broadening parameter (systematics), which defines the
width of the lithium lines. This parameter, due to stellar rotation
and macro-turbulent motions in the atmosphere, is constrained from the
profile fitting of other lines, such as Fe{\sc i} and Ca{\sc i}
\cite{nissen}. Therefore, in order to reach very low $^6$Li/$^7$Li
ratios in metal-poor stars, one has to considerably diminish the
statistical noise, and, at the same time, to carefully control
systematics.

Concerning the statistical accuracy, we note, as a reference, that the
most precise measurement of the $^6$Li/$^7$Li ratio was carried out in
the star HD84937, of magnitude m$_V=8.3$, in 1hr. integration time on
the 2.7m McDonald Telescope and Coud\'e Spectrometer, at a resolving
power $\lambda/\Delta\lambda=1.25\times10^5$, yielding
$^6$Li/$^7$Li=5\%$\pm$2\% (statistical and systematics combined)
\cite{nissen,hobbs,nissen2,nissen3}.  The noise could be reduced by a
factor of $\simeq6$ for an integration time $\sim$20hrs. on an
instrument such as the 3.9m Anglo-Australian Telescope, assuming equal
efficiencies for the spectrometers. On future telescopes such as one
8m reflector at the European Southern Observatory Very Large
Telescope, using the UVES spectrograph, this factor could be brought
up to $\simeq12$ for 20hrs. integration time. However, this factor
would not compensate for the difference of magnitude for a star at
very low metallicities, since a factor of 12 allows one to achieve the
same signal-to-noise ratio on a star 2.7 magnitudes higher. Indeed,
HD84937 is a uniquely bright target; in the very metal-poor star
survey by Beers, Preston \& Shectman \cite{beers,beers3,beers2}, we do
not find any star brighter than $m_V=13$, for [Fe/H]$<-3$ and a
temperature of about $T_{eff}>$6000 K, needed to ensure that $^6$Li
has not been depleted ( {\it i.e.} destroyed and/or diluted) too much.
This survey is not complete yet, and
one may still hope to find a suitable candidate. At the present time,
however, the prospect of detecting $^6$Li at a level
$^6$Li/$^7$Li$<1$\% does not seem realistic at metallicities
[Fe/H]$<-3$. Only at the high values of $^6$Li/$^7$Li $\leq 3$\%
allowed by the present experimental upper limit on deuterium-alpha
capture of \cite{cecil}, could even undepleted primeval $^6$Li be
detected with present instruments.

Regarding the systematics, it is unfortunately difficult to evaluate to 
what level these errors could be brought down. One would clearly have to 
increase the number of profiles studied to determine more accurately the 
theoretical line profile. In that frame, increasing the resolving 
power up to $\lambda/\Delta\lambda\sim3.\times10^5$ would 
help considerably,  although an increase in spectral resolution is
associated with a lower signal-to-noise ratio per resolution element.

\section{Conclusion}

We examined possible $^{6}{\rm Li}$ abundances predicted by Big-Bang
Nucleosynthesis, and discussed the uncertainties in these predictions.
The latter arise primarily from the uncertainties in the rate of the
$^{2}{\rm H}(\alpha,\gamma)^{6}{\rm Li}$ radiative capture reaction,
which determines the final yield of $^6$Li. These uncertainties arise
because this cross-section has never been measured directly at the
relevant energies for Big-Bang production of $^6$Li, where the
cross-section falls steeply with decreasing energy.  Uncertainties in
theoretical estimates amount to roughly a factor 10 on the yield of
$^6$Li, and, as such, would preclude putting severe constraints on the
baryonic density parameter $\Omega_B$ from $^6$Li alone, if a primeval
component of $^6$Li were observed.  The experimental upper limit of
Cecil {\it et al.} \cite{cecil} also allows the $^6$Li yield to be
considerably higher than allowed by any of these estimates, so that
any constraint on $\Omega_B$ from $^6$Li alone would be difficult to
arrive at.  However, since significant $^6$Li yields are favored by
low baryon density and are strongly suppressed at high baryon density,
regardless of the possible value of the production cross section, any
detection of primordial $^6$Li would favor the low end of the current
$\Omega_B$ range from BBN.  This would favor higher primordial $D/H$
values.  Thus, we emphasize that the detection of any
primordial $^6$Li, to a level $\log$($^6$Li/H)$\sim-14\to-12$, as
obtained from our present calculations, would provide a new
fundamental test of Big-Bang nucleosynthesis, hence of modern
cosmology, and it could help resolve the current debate over which
value of the extragalactic deuterium-to-hydrogen ratio is
representative of the primordial value.

Finally, we caution that the prospect of detecting $^6$Li in the
atmospheric layers of a very metal-deficient star (pristine material),
appears marginal with current instrumentation. With the present
instruments available, and even for the larger instruments currently
under construction, it seems that primordial $^6$Li could be detected,
in stars with [Fe/H]$<-3$, only near the extreme upper error bar of
the $d + \alpha$ reaction rate and the low end of the allowed baryon
density, for which $\log$($^6$Li/H)$\sim-12$.  Clearly, measurements
of the $d + \alpha$ cross section at relevant energies are crucial for
deciding whether or not observational techniques should be pushed in
this direction.

\section*{Acknowledgements}

We would like to thank B. Fields, R. Wiringa and R. G. H. Robertson
for useful information, as well as D. Thomas for providing code for
the extended reaction network.  This work was supported in part by the
DOE, NASA, and the NSF at the University of Chicago, and by the DOE
and NASA at Fermilab.

\newpage

\widetext
\begin{table}
\squeezetable
\caption{Reaction rates that determine the primordial $^6$Li
abundance, roughly in order of importance.  Errors have only been
assessed for reactions determined to affect the final $^6$Li
abundance significantly.}
\begin{tabular}{lr@{}lcll}
Reaction &\multicolumn{3}{c}{Rate, $\langle N_{A}\sigma v
\rangle$  (cm$^3$s$^{-1}$)} &  $2\sigma$ Error  & Source\\
\tableline
$^{2}{\rm H}(\alpha,\gamma)^{6}{\rm Li}$ &  1.79 & $\times10^{3} 
T^{-2/3}_{9}\exp(-7.429/T_{9}^{1/3}) (1+.056T_{9}^{1/3})$
 & $T_{9}< 3.1$ & Upper &
Present Work\\
&& $+ 9.71\times10^{1}T_{9}^{-3/2}\exp(-8.251/T_{9})$ & & Limit\\
& 3.01 & $\times 10^{1}T_{9}^{-2/3}\exp(-7.423/T_{9}^{1/3})$ &
$T_{9} > 3.2$ \\
&&\multicolumn{2}{l}{$\times(1.+.056
T_{9}^{1/3}-4.85T_{9}^{2/3}+8.85T_{9}-.585T_{9}^{4/3}-.584T_{9}^{5/3})$}\\
&& $+ 9.71\times 10^{1}T_{9}^{-3/2}\exp(-8.251 /T_{9})$\\
$^{6}{\rm Li}(p,\alpha)^{3}{\rm He}$ & 3.39 & $\times10^{10} T_{9}^{-2/3}
{\rm exp}(-8.415/T_{9}^{1/3} - (T_{9}/5.50)^{2})$ && 15\%
& Present work 
\\ 
& &\multicolumn{2}{l}{$\times(1 + 0.0495T_{9}^{1/3} - 0.087T_{9}^{2/3} - 0.030T_{9} 
- 0.0055T_{9}^{4/3} - 0.0048T_{9}^{5/3})$}\\
& &\multicolumn{2}{l}{$+1.33\times 10^{10}T_{9}^{-3/2}\exp(-17.793/T_{9})$}\\
& &\multicolumn{2}{l}{$+1.29\times 10^{9}T_{9}^{-1}\exp(-21.820/T_{9})$}\\
$^{6}{\rm Li}(n,\alpha)^{3}{\rm H}$ & 2.54 & $\times 10^{9}
T_{9}^{-3/2} \exp(-2.39/T_{9})$ && --- & Caughlan and Fowler\tablenotemark[1]\\
$^{3}{\rm He}(t,\gamma)^{6}{\rm Li}$ & 2.21 & $\times 10^{5}
T_{9}^{-2/3}\exp(-7.720/T_{9}^{1/3})$ && --- & Fukugita and
Kajino\tablenotemark[2]\\
&& \multicolumn{2}{l}{$\times (1 + 2.68T_{9}^{2/3} + 0.868T_{9} +
0.192T_{9}^{4/3} + 0.174T_{9}^{5/3} + 0.044T_{9}^{2})$}\\
$^{6}{\rm Li}(n,\gamma)^{7}{\rm Li}$ & 5.10 & $\times 10^{-3}$ &&
---  & Malaney and Fowler\tablenotemark[3]\\
$^{4}{\rm He}(nn,\gamma)^{6}{\rm He}$ &4.04 & 
\multicolumn{2}{l}{$\times10^{-12}T_{9}^{-2}\exp(-9.585/T_{9})
(1 + 0.138T_{9})$} & --- & Caughlan and
Fowler\\
$^{6}{\rm He} \rightarrow e $+$ ^{6}{\rm Li}$ & 0.85&9 && --- & Malaney
and Fowler \\
$^{6}{\rm Li}(p,\gamma)^{7}{\rm Be}$ & 6.69 & \multicolumn{2}{l}{$\times 10^{5}
T_{9a}^{5/6}T_{9}^{-3/2}\exp(-8.413/T_{9}^{1/3})$} & --- & 
Caughlan and Fowler\\
$^{9}{\rm Be}(p,\alpha)^{6}{\rm Li}$ & 2.11 & \multicolumn{2}{l}
{$\times 10^{11}T_{9}^{-2/3}\exp(-10.359/T_{9}^{1/3} - (T_{9}/0.520)^{2})$} 
& --- & Caughlan and Fowler \\
&&\multicolumn{2}{l}{$\times (1 + 0.040T_{9}^{1/3} + 1.09T_{9}^{2/3}
+ 0.307T_{9} + 3.21T_{9}^{4/3} + 2.3T_{9}^{5/3})$}\\
&&\multicolumn{2}{l}{$ +4.51\times 10^{8}T_{9}^{-1}\exp (-3.046/T_{9})$}\\
&&\multicolumn{2}{l}{$ +6.70\times 10^{8}T_{9}^{-3/4}\exp (-5.16/T_{9})$}\\
$^{6}{\rm Li}(\alpha,\gamma)^{10}{\rm B}$ & 4.06 & \multicolumn{2}{l}
{$ \times 10^{6}T_{9}^{-2/3}\exp(-18.790/T_{9}^{1/3} - (T_{9}/1.326)^{2})$} &
--- & Caughlan and Fowler \\
&& \multicolumn{2}{l}{$\times (1 + 0.22T_{9}^{1/3} + 1.54T_{9}^{2/3} 
+ 0.239T_{9} + 2.20 T_{9}^{4/3} + 0.869T_{9}^{5/3})$}\\
&&\multicolumn{2}{l}{$ +1.91\times 10^{3}T_{9}^{-3/2}\exp (-3.484/T_{9})$}\\
&&\multicolumn{2}{l}{$ +1.01\times 10^{4}T_{9}^{-1}\exp (-7.269/T_{9})$}\\
$^{6}{\rm Li}(d,n)^{7}{\rm Be}$ & 1.48 & \multicolumn{2}{l}
{$\times 10^{12}T_{9}^{-2/3}\exp(-10.135/T_{9}^{1/3})$} & --- & 
Malaney and Fowler\\
$^{6}{\rm Li}(d,p)^{7}{\rm Li}$ & 1.48 & \multicolumn{2}{l}
{$\times 10^{12}T_{9}^{-2/3}\exp(-10.135/T_{9}^{1/3})$} & --- & 
Malaney and Fowler\\
$^{9}{\rm Li}(p,\alpha)^{6}{\rm He}$ & 1.03 & \multicolumn{2}{l}
{$\times 10^{11}T_{9}^{-2/3}\exp(-8.533/T_{9}^{1/3})$} & --- &
Thomas {\it et al.}\tablenotemark[4]\\
\end{tabular}
\tablenotetext[1]{Ref. \protect\cite{CF}.}
\tablenotetext[2]{Ref. \protect\cite{fukugita}.}
\tablenotetext[3]{Ref. \protect\cite{mf}.}
\tablenotetext[3]{Ref. \protect\cite{thomas}.}
\label{table:rates}
\end{table}


\end{document}